\documentclass[pra,twocolumn,superscriptaddress,10pt,noshowpacs]{revtex4}
\usepackage[english]{babel}
\usepackage[T1]{fontenc}
\usepackage[utf8]{inputenc}
\usepackage{graphicx,epstopdf}
\usepackage{amsmath}

\usepackage[pdftex,plainpages=false,colorlinks=true,citecolor=blue,linkcolor=blue,urlcolor=blue,filecolor=green,bookmarksopen=true]{hyperref}
\usepackage{amsfonts}
\usepackage{bbm}
\usepackage{amssymb}
\usepackage{color}
\usepackage{latexsym}
\usepackage{times,txfonts}

\begin{document}

\title{Anisotropic Ginzburg-Landau model for superconductivity with five-dimensional operators}

\author{M. C. Ara\'{u}jo}
\email{michelangelo@fisica.ufc.br}
\affiliation{Universidade Federal do Cear\'a (UFC), Departamento de F\'isica,\\ Campus do Pici, Fortaleza - CE, C.P. 6030, 60455-760 - Brazil.}

\author{I. C. Jardim}
\affiliation{Departamento de F\'{i}sica, Universidade Regional do Cariri, 57072-270, Juazeiro do Norte, Cear\'{a}, Brazil}

\author{D. F. S. Veras}
\affiliation{Centro de Ci\^{e}ncias e Tecnologia, Universidade Federal do Cariri, 63048-080, Juazeiro do Norte, Cear\'{a}, Brazil}

\author{J. Furtado}
\affiliation{Centro de Ci\^{e}ncias e Tecnologia, Universidade Federal do Cariri, 63048-080, Juazeiro do Norte, Cear\'{a}, Brazil}
\affiliation{Department of Physics, Faculty of Science, Gazi University, 06500 Ankara, Turkey}
\email{job.furtado@ufca.edu.br}

\date{\today}

\begin{abstract}
This paper presents the effects of non-minimal Lorentz-violation operators in superconductivity. By constructing a Lorentz-Violating Ginzburg-Landau theory of superconductivity with a five-dimensional operator, we discuss the influence of higher dimensional Lorentz-Violating operators in the London's depth penetration, in the coherence length and critical magnetic field.  
\end{abstract}

\maketitle

\section{Introduction}

In recent years, there has been a focus on exploring potential expansions of the Standard Model (SM). Within this framework, the discussion has shifted towards the breaking of Lorentz and CPT symmetries, considered as a significant subject \cite{Kos1, Kos2, Kos3, Kos4, Colladay:1996iz, Colladay:1998fq}. Typically, Lorentz symmetry is disrupted by introducing specific directions in space-time, represented by additive terms that are proportionate to small constant vectors or tensors. The Standard Model Extension (SME) \cite{Colladay:1996iz, Colladay:1998fq} stands out as the most established model addressing the consequences of Lorentz and CPT symmetry violation. SME, being an effective field theory, incorporates all conceivable terms violating Lorentz and CPT symmetries in its Lagrangian.

Kostelecky \cite{Edwards:2018lsn}. proposed an extension of the scalar sector incorporating Lorentz-violating effects. This model presents a comprehensive effective scalar field theory across any spacetime dimension, featuring explicit perturbative spin-independent Lorentz-violating operators of arbitrary mass dimension. The significance of this development lies in the predominance of spin in most fundamental particles of the SM, with the Higgs boson being the sole example of a fundamental spinless particle. Despite the relatively minor role of the scalar sector of QED (sQED) compared to strong interactions in describing meson coupling, it has been suggested \cite{Edwards:2019lfb} that a Lorentz-violating extension of sQED could effectively address slight CPT deviations in neutral-meson oscillations.

The minimal extension of SME was extensively studied in the last years in several contexts, such as radiative corrections \cite{Jackiw:1999yp, Kostelecky:2001jc, Ferrari:2021eam}, neutrinos oscillations \cite{MINOS:2008fnv, MINOS:2010kat, MiniBooNE:2011pix}, Euler-Heisenberg effective action \cite{Ferrari:2021eam, Furtado:2014cja}, gravitational context \cite{Kostelecky:2016kkn, Assuncao:2018jkq}, finite temperature field theory \cite{Assuncao:2018jkq, Leite:2013pca, Leite:2011jg, Assuncao:2016fko, Araujo:2023izx} among others (for a more complete review, see f.e. \cite{Mariz:2022oib}). The non-minimal extension, despite its posses only nonrenormalizable terms, have been receiving attention in the literature \cite{Myers:2003fd, Kostelecky:2009zp, Kostelecky:2011gq, Mariz:2010fm, Rubtsov:2012kb}. A strong motivation lies in the fact that some relevant astrophysical processes impose severe restrictions on the coefficients associated with the operators with dimension $d \geq 5$, being such contributions comparable or even dominant when compared with the ones that arise from dimension $d \leq 4$ operators. Within the nonminimal LV framework, an important role was played by the paper \cite{Kostelecky:2009zp}, in which the simplest cases of such operators were, for the very first time, introduced in the LV scenario. The first studies of their perturbative impact were investigated in \cite{Gomes:2009ch, Mariz:2010fm, Gazzola:2010vr, Mariz:2011ed, BaetaScarpelli:2013rmt}. And finally, all possible LV extensions of SME with dimensions up to 6 for fermion-dependent operators were listed in \cite{Kostelecky:2018yfa}.

The potential applications of models violating Lorentz and CPT symmetries span various fields, extending from quantum field theory to condensed matter physics. The intersection of Lorentz-violating models and condensed matter physics has garnered significant attention, particularly in the context of Weyl semimetals \cite{Assuncao:2015lfa, Gos1, Gos2, Gos3}, superconductivity \cite{Bazeia:2016pra, Furtado:2020iom, Granado:2023czl}, graphene with anisotropic scaling \cite{Katsnelson:2012cz}, dark matter and black holes analog models \cite{Baym:2020uos, Pereira:2009vb}, among others.

The Ginzburg-Landau (GL) theory of superconductivity serves as a construct within effective quantum field theory, elucidating several crucial facets of superconductivity \cite{Ginzburg:1950sr}. Despite its initial classification as a phenomenological theory, the GL theory can be construed as a specific case arising from the Bardeen-Cooper-Schrieffer (BCS) theory of superconductivity \cite{Bardeen:1957mv}. The exploration of superconductivity in diverse realms, such as Weyl semimetals \cite{Zhou:2015qka, Bednik:2015tha, Wei:2014vsa} and carbon-based nanostructures like graphene and fullerene \cite{Roy:2013aya, Cohnitz:2017vsr} is frequently undertaken in literature due to their potential applications for technological advancements. However, recent years have seen an intensified discourse on the theoretical aspects of superconductivity, extending into domains such as astrophysics \cite{Madsen:1999ci, Bonanno:2011ch}, cosmology \cite{Ebert:2007ey, Gao:2012aw} and in high energy physics \cite{Herzog:2009xv, Ghoroku:2019trx}. Recently, studies of anisotropic superconductors were carried out by considering Lorentz-violating terms inspired by the SME \cite{Furtado:2020iom}. The authors have shown in \cite{Furtado:2020iom} that the coherence length and critical magnetic field are not affected by Lorentz-violating modifications in the scalar sector, but the London's depth penetration in modified by the Lorentz-violating contributions.

In this study we consider a Lorentz-violating complex scalar sector coupled non-minimally with the gauge field, the usual Maxwell term as the gauge sector and $\lambda|\phi|^4$ promoting the spontaneous symmetry breaking. We discuss the influence of a higher dimensional Lorentz-Violating operator in the London's depth penetration, in the coherence length and critical magnetic field. The Lorentz-violating parameters can be interpreted as defects or layers in the superconductor, giving rise to an anisotropy in the system.

This paper is organized as follows: in the next section we present our model, discussing the discrete symmetries and the non-minimal Lorentz-violating extension of the Ginzburg-Landau theory for superconductivity. In the section III we discuss the superconducting case for the Lorentz-violating contributions to the coherence length, critical magnetic field and London's depth penetration. Finally, in section IV we highlight our conclusions.

\section{Model}
The model we are considering consists of the complex scalar sector coupled non-minimally with the gauge field, the usual Maxwell term as the gauge sector and $\lambda|\phi|^4$ potential promoting the spontaneous symmetry breaking. Hence the Lagrangian describing the system is 
\begin{eqnarray}\label{lagrangian1}
\nonumber\mathcal{L}&=&(\nabla_{\mu}\phi)^{*}(\nabla^{\mu}\phi)-m^2\phi^{*}\phi-\lambda(\phi^{*}\phi)^2-\frac{1}{4}F^{\mu\nu}F_{\mu\nu},
\end{eqnarray}
where 
\begin{eqnarray}\label{covderivnominimal1111}
\nabla_{\mu}\phi= \left( D_{\mu} -i\frac{g}{2}\epsilon_{\mu\nu\alpha\beta}\omega^{\nu}F^{\alpha\beta}\right) \phi
\end{eqnarray}
is the non-minimal covariant derivative and $F_{\mu\nu}=\partial_{\mu}A_{\nu}-\partial_{\nu}A_{\mu}$. The usual covariant derivative in turn is given by $D_{\mu}=\partial_{\mu}-ieA_{\mu}$. In the high-energy physics scenario, the vector $\omega^{\mu}$ promotes the violation of Lorentz invariance by setting a privileged direction in space-time and breaking the equivalence between particle and observer transformations. Assuming that the vector is constant, it implies the independence of space-time position, which yields translational invariance and assures the conservation of momentum and energy. The interest in this specific model of Lorentz symmetry violation lies in the fact that the covariant derivative in Eq. \eqref{covderivnominimal1111} contains a Pauli-type coupling term, where the complex scalar field directly couples to the electromagnetic field  strength. Notice that the product $g\omega^{\mu}$ must have dimension of $(\text{mass})^{-1}$ and it plays the role of a Lorentz violating induced magnetic moment. In $(2+1)$ dimensions, it has been shown that a similar non-minimal coupling term gives rise to fractional spin statistics even in the absence of the Chern-Simons term \cite{Nobre:1999mj, Furtado:2011zz}, which is known to exhibit such a characteristic \cite{Semenoff:1988jr,Dunne:1994uy}. Fractional statistics play an important role in the interaction between quantum field theory and condensed matter physics \cite{Nobre:1999mj,Wilczek:1982wy}.
 
The kinetic term can be written as
\begin{eqnarray}
\nonumber(\nabla_{\mu}\phi)^{*}(\nabla^{\mu}\phi)&=&(D_{\mu}\phi)^{*}(D^{\mu}\phi)\\
\nonumber&+&i\frac{g}{2}\epsilon_{\mu\nu\alpha\beta}\omega^{\nu}F^{\alpha\beta}\left[\phi^{*}(D^{\mu}\phi)-(D^{\mu}\phi)^{*}\phi\right]+\cdots,\\
\end{eqnarray}
where the dots stands for higher-orders Lorentz breaking terms.

In the static condition the Lagrangian takes the form:
\begin{eqnarray}\label{lagrangian2}
    \nonumber\mathcal{L}&=& \eta^{ij}(D_i\phi)^*(D_j\phi)-\mu^2\phi^*\phi-\lambda(\phi^*\phi)^2-\frac{1}{4}F_{ij}F^{ij}\\
    &-& i\, \frac{g}{2}W\epsilon_{ijk}F^{jk}\left[\phi^{*}D^i\phi-(D^i\phi)^{*}\phi\right].
\end{eqnarray} Now, $-\mathcal{L}$ is the anisotropic Ginzburg-Landau free energy with a five-dimensional operator. Note that the mass parameter $m^2$ was written as $\mu^2$, which is now considered a temperature-dependent system parameter near the critical temperature $T=T_c$. Specifically,  $\mu^2=a\, (T-T_c)$. We have further identified the temporal component of the vector $\omega^{\mu}$ as being $W$, i.e., $W=\omega^{0}$. In this context, $\phi$ is the macroscopic many-particle wave function. The physical interpretation of $\phi$ as a many-particle wave function is justified by the Bardeen-Cooper-Schrieffer (BCS) theory, according to which, under certain conditions, there is an attractive force between electrons, and field quanta are electron pairs, which are, of course, bosons. At low temperatures, the field quanta fall into the same quantum state (Bose-Einstein Condensation \cite{Furtado:2020olp, Casana:2011bv}), and because of this, a many-particle wave function $\phi$ may be used to describe the macroscopic system. In this sense, the quantity $e$ in the usual covariant derivative expression should be understood as being twice the value of the electric charge.

The equation of motion for the gauge field is given by
\begin{eqnarray}
  \mathbf{\nabla}^2 A^{m} + \partial_n\partial^mA^n = - ej^m,  
\end{eqnarray}
where
\begin{eqnarray}\label{densidadedecorrenteconserv}
    j^m &=& -i\phi^{*}\overleftrightarrow{\partial^m}\phi-2eA^m\phi^{*}\phi+gW(\mathbf{x})\epsilon^{mjk}F_{jk}\phi^{*}\phi \nonumber\\
    &+& i\, \frac{g}{e}\partial_nW(\mathbf{x})\epsilon^{imn}(\phi^{*}\overleftrightarrow{\partial_i}\phi-2ieA_i\phi^{*}\phi)\nonumber\\
    &+& i\, \frac{g}{e}W(\mathbf{x})\epsilon^{imn}\Bigl[ 2\partial_n\phi^{*}\partial_i\phi + \phi^{*}\overleftrightarrow{\partial_{ni}}\phi \nonumber\\
    &-& 2ie\partial_nA_i\phi^{*}\phi - 2ieA_i( \phi^{*}\partial_n\phi + \partial_n\phi^{*}\phi ) \Bigr] 
\end{eqnarray} is the conserved current density due to the invariance of the theory under global $U(1)$ symmetry. In equation \eqref{densidadedecorrenteconserv},
\begin{eqnarray}
  \phi^{*}\overleftrightarrow{\partial_{i_1,\cdots, i_n}}\phi = \phi^{*}(\partial_{i_1}\cdots\partial_{i_n})\phi - (\partial_{i_1}\cdots\partial_{i_n})\phi^{*}\phi,   
\end{eqnarray} and we have taken into account the possibility that the Lorentz violation parameter is a function that depends on position, i.e. $W=W(\mathbf{x})$.

% \begin{eqnarray}
%     \nonumber&&ie(\phi^*\partial^m\phi-\phi\partial^m\phi^*)+2e^2A^m|\phi|^2+\Box A^m-\partial_k\partial^mA^k\\
%    \nonumber&&-ig[\partial_nW(x)]\epsilon_{i}^{nm}\left(\phi\partial^{i}\phi^{*}-\phi^{*}\partial^{i}\phi+2ieA^{i}|\phi|^2\right)\\
%     \nonumber&&-geW(x)\epsilon^m_{jk}F^{jk}|\phi|^2-igW(x)\epsilon_{i}^{nm}\times\\
%      \nonumber&&\left[2\partial_n\phi\partial^{i}\phi^{*}+2ie(\partial_n\phi)A^{i}\phi^{*}+\phi\partial_n\partial^{i}\phi^{*}\right.\\
%     &&\left.2ie(\partial_nA^{i})|\phi|^2+2ie(\partial_n\phi^{*})A^{i}\phi-\phi^{*}\partial_n\partial^{i}\phi\right]=0,
% \end{eqnarray}
% allow us to obtain the conserved current for the model, which reads as
% \begin{eqnarray}
%     \nonumber j^{i}(x)&=&-i(\phi^*\partial^{m}\phi-\phi\partial^{m}\phi^*)-2eA^m|\phi|^2-geW(x)\epsilon^m_{jk}F^{jk}|\phi|^2\\
%     \nonumber&&-ig[\partial_nW(x)]\epsilon_{i}^{nm}\left(\phi\partial^{i}\phi^{*}-\phi^{*}\partial^{i}\phi+2ieA^{i}|\phi|^2\right)\\
%     \nonumber&&-igW(x)\epsilon_{i}^{nm}\left[2\partial_n\phi\partial^{i}\phi^{*}+2ie(\partial_n\phi)A^{i}\phi^{*}+\phi\partial_n\partial^{i}\phi^{*}\right.\\
%     &&\left.2ie(\partial_nA^{i})|\phi|^2+2ie(\partial_n\phi^{*})A^{i}\phi-\phi^{*}\partial_n\partial^{i}\phi\right].
% \end{eqnarray} 
 % Notice that we have considered the possibility of a position-dependent $W(x)$ function. Such contributions will be addressed independently in what follows. 
 Considering that $\phi$ varies only very slightly over the sample, so that $\partial_i\phi\approx 0$, we can simplify the current density $j^{m}=j^{m}(\mathbf{x})$ to
 \begin{eqnarray}\label{jmxvectormodphi27}
     j^m(\mathbf{x}) &=& |\phi|^2 \Bigl[ -2eA^m+gW(\mathbf{x})\epsilon^{mjk}F_{jk}\nonumber\\ &+&2g\epsilon^{imn}\partial_nW(\mathbf{x})A_i+2g\epsilon^{imn}W(\mathbf{x})\partial_nA_i\Bigr]
 \end{eqnarray}
% \begin{eqnarray}
%     \nonumber j^{i}(x)&=&-2eA^m|\phi|^2-geW(x)\epsilon^m_{jk}F^{jk}|\phi|^2+2ge[\partial_nW(x)]\epsilon_{i}^{nm}2A^{i}|\phi|^2\\
%     &&+2geW(x)\epsilon_{i}^{nm}(\partial_nA^{i})|\phi|^2.
% \end{eqnarray}

% The above conserved current is in connection with the $U(1)$ symmetry of the lagrangian, so that, according to Noether's theorem, for any given continuous symmetry there is an associated conserved current.

Now, the ground state where Bose-Einstein Condensation is conceived can be obtained by minimizing the potential $V(\phi^*,\phi)$ for a temperature $T<T_c$, i.e., $\mu^2<0$. Since 
\begin{equation}
V(\phi^*,\phi)=\mu^2|\phi|^2+\lambda|\phi|^4,
\end{equation}  the free energy minimum occurs at 
\begin{equation}
    |\phi|^2=-\frac{\mu^2}{2\lambda}>0,
\end{equation}
where, by definition, $\lambda >0$. In this configuration, the current density in Eq. \eqref{jmxvectormodphi27} takes the form of
\begin{eqnarray}\label{jvectorxgamma2violation}
    \mathbf{j}(\mathbf{x}) = -\Gamma^2\mathbf{A} - \frac{g}{e}\Gamma^2\Biggl[ 2W(\mathbf{x})(\mathbf{\nabla}\times\mathbf{A})+\mathbf{\nabla}W(\mathbf{x})\times\mathbf{A} \Biggr]
\end{eqnarray}
% \begin{eqnarray}
% \vec{j}=-\Gamma^2\vec{A}-2g\Gamma^2W(x)\left(\nabla\times\vec{A}\right)+g\Gamma^2\left(\vec{A}\times\nabla W(x)\right).
% \end{eqnarray}

where
\begin{eqnarray}
\Gamma^2\equiv-\frac{e\mu^2}{\lambda},
\end{eqnarray} with $\Gamma$ being a positive definite constant. From now on, we will only work with the case when $T<T_c$. Therefore, no phase transitions will be considered. Eq. \eqref{jvectorxgamma2violation} is the so called London equation, modified by the presence of the Lorentz-violating function $W(\mathbf{x})$. 

Now, Ohm’s law defines resistance $R$ through the equation $\mathbf{E} = R\, \mathbf{j}$. In the static condition of our model, the electric field is simply $\mathbf{E}=-\partial\mathbf{A}/\partial t = \mathbf{0}$. As a result, we have two possible solutions: $R=0$ (superconducting case) or $\mathbf{j}=\mathbf{0}$ (insulating case).

A non-trivial solution for the insulating case occurs when $W(\mathbf{x})\neq 0$. To ensure that the current is zero within the insulating material, the vector potential and the Lorentz violation parameter must satisfy 
\begin{eqnarray}\label{vetorpotencialcomviolac}
  \mathbf{A} + \frac{g}{e}\, \Biggl[ 2W(\mathbf{x})(\mathbf{\nabla}\times\mathbf{A})+\mathbf{\nabla}W(\mathbf{x})\times\mathbf{A} \Biggr] = \mathbf{0}.
\end{eqnarray} From Faraday’s law we know that the magnetic field must also be static and since Ampère-Maxwell equation leads to $\mathbf{\nabla}\times\mathbf{B}=\mathbf{0}$, taking the curl from Eq. \eqref{vetorpotencialcomviolac}, we find
\begin{eqnarray}\label{magbcomviolactemosdepotvec}
     2\, \mathbf{B}\times\mathbf{\nabla}W(\mathbf{x}) - \frac{e}{g}\, \mathbf{B}   &=& \bigl(\mathbf{A}\cdot\mathbf{\nabla}\bigr)\, \mathbf{\nabla}W(\mathbf{x}) - \bigl(\mathbf{\nabla}W(\mathbf{x})\times \mathbf{\nabla}\bigr)\, \mathbf{A} \nonumber\\
    &+& \mathbf{\nabla}W(\mathbf{x})\, (\mathbf{\nabla}\cdot\mathbf{A}) - \mathbf{A}\, \mathbf{\nabla}^2W(\mathbf{x}).
\end{eqnarray} As we can notice from Eq. \eqref{magbcomviolactemosdepotvec}, the magnetic field will be zero if the violation parameter remains constant over the material.

\section{Superconducting case}

In this section we will study the superconducting case ($R=0$) for both constant and position-dependent Lorentz-violating parameter. In the context of GL theory of superconductivity, three important properties can be obtained, namely, the coherence length, the critical magnetic field, and London's penetration depth. Let’s investigate each of these properties separately in the following.

\subsection{The coherence length}

In order to obtain the coherence length of the superconductor we have to solve the equation of motion for the complex scalar field, which is given by
\begin{eqnarray}\label{eqdifescalarcomplexequatio}
    \Lambda\phi-2\lambda\, |\phi|^2\phi &=& ig\, \Bigl\{2W(\mathbf{x})\, \bigl[ (\mathbf{\nabla}\times\mathbf{A})\cdot\mathbf{D}\bigr] \nonumber\\
    &-& \bigl[(\mathbf{\nabla}\times\mathbf{A})\cdot\mathbf{\nabla}W(\mathbf{x})\bigr]\Bigr\}\phi
\end{eqnarray} where we have defined $\Lambda$ as 
\begin{eqnarray}
    \Lambda \equiv \mathbf{\nabla}^2-\mu^2+2ie\, \mathbf{A}\cdot\mathbf{\nabla}+ie\, \mathbf{\nabla}\cdot\mathbf{A}-e^2\, \mathbf{A}^2,
\end{eqnarray} and used the fact that $\mathbf{D} = -\mathbf{\nabla} -ie\mathbf{A}$. However, the coherence length is a property that characterizes superconducting materials and is independent of the existence of an external magnetic field. For that reason we can assume $\mathbf{A}=\mathbf{0}$, since $\mathbf{A}$ is the vector potential that arises within the material in response to the external magnetic field. Regarding boundary conditions, we usually desire the field $\phi$ to be zero at the surface of the superconductor and acquire a nearly constant maximum value of $\phi_{max}$ deep inside the material. Note from Eq. \eqref{eqdifescalarcomplexequatio} that $|\phi_{max}|^2=-\mu^2/2\lambda$. Based on what was discussed above, it is straightforward to show that the superconducting electrons density must be given by
\begin{eqnarray}
    |\phi|^2=|\phi_{max}|^2\tanh^2\left( \frac{x}{\sqrt{2}\, \xi} \right)
\end{eqnarray} for a semi-infinite superconductor occupying the space $x>0$. The parameter $\xi$, given by
\begin{eqnarray}
    \xi = \sqrt{\frac{1}{|\, \mu^2\, |}},
\end{eqnarray} defines the coherence length. Note that this parameter is insensitive to Lorentz-violating effects.

% In order to obtain the coherence length of the superconductor we have to solve the equation of motion for $\phi$ considering the following boundary conditions: in the surface of the superconductor we must have $\phi=0$ and deep inside the material $\phi=\phi_{max}$, so that $\phi_{max}$ is practically constant, which means that $\partial_i\phi_{max}=0$. Hence, the highest density of superconducting electrons occurs as usual at $|\phi_{max}|^2=-2\mu^2/\lambda$. Thence, it is straightforward to find that
% \begin{equation}
%     |\phi|^2=|\phi_{max}|^2\tanh^2\left({\frac{\sqrt{2}x}{2\xi}}\right),
% \end{equation}
% with the coherence length $\xi$ being given by
% \begin{equation}
%     \xi=\sqrt{\frac{\hbar^2}{2m^*|\mu^2|}},
% \end{equation}
% where $m^*=2m_e$, being $m_e$ the effective electron's mass. The factor of 2 takes into account the fact that the $\phi$ field stands for Cooper pairs. 

%\begin{figure}[h!]
%\begin{center}
%\includegraphics[scale=0.4]{SCfig2.eps}
%\caption{Coherence length for the LV-superconductivity}
%\label{fig2}
%\end{center}
%\end{figure}

\subsection{The critical magnetic field}

The critical magnetic field is defined as the minimum value of the external magnetic field $H$ above which the state of superconductivity is lost. For a type-I superconductor, the critical magnetic field $H_c$ is directly calculated from the Ginzburg-Landau free energy under the condition that it is the maximum value of the external magnetic field for which the free energy of the superconducting state is less than that of the normal state.  In this sense, in the presence of an external magnetic field and at the limit where superconductivity is maximum, we can write
\begin{eqnarray}
    \mu^2\, |\phi_{max}|^2 + \lambda\, |\phi_{max}|^4 + \mu_0\, \frac{H_c^2}{2} = 0
\end{eqnarray} so that
\begin{eqnarray}\label{campmagcriticosupconduct1}
    H_c=\frac{|\mu^2|}{\sqrt{2\lambda \mu_0}}.
\end{eqnarray} Note from Eq. \eqref{campmagcriticosupconduct1} that the critical magnetic field for a type-I superconductor is not affected by the presence of Lorentz violation in our model.

\begin{figure}
\includegraphics[scale=0.74]{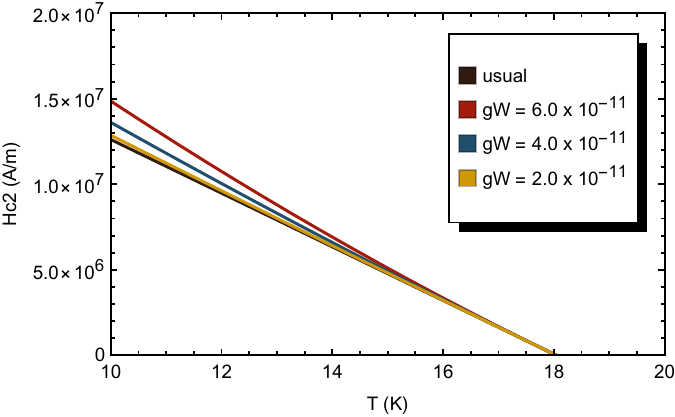}
    \caption{Graph of the critical magnetic field $H_{c2}$ as a function of the temperature $T$ for different values of the product $gW$. We have considered $a=6.3 \times 10^{-19}$ and $T_c = 18.05\, K $ (Critical temperature of the type II superconductor $Nb_3Sn$).}
    \label{hc2xtgraph}
\end{figure}

In type-II superconductors, the value of $H_c$ is not sufficient to break the superconducting state, since the material can exist in a vortex state. To determine the critical magnetic field $H_{c2}$ above which the normal state of a type-II superconducting material is reached, we must analyze the equation of motion for the complex scalar field in a specific geometry. Let us consider a semi-infinite type-II superconductor occupying the space $x>0$ under an external magnetic field $\mathbf{H} = H\hat{\mathbf{z}}$ with $H$ constant. Since magnetization is small, especially in strongly type-II superconductors, we can assume that inside the superconductor $B \approx \mu_0 H$. Therefore, we can suppose a vector potential of the form
\begin{eqnarray}
  \mathbf{A} = x B \hat{\mathbf{y}},  
\end{eqnarray} which is consistent with the Landau gauge ($\mathbf{\nabla}\cdot \mathbf{A} =0$). Also, for the sake of simplicity, we will assume that the Lorentz violation parameter is a constant $W$. Thus, since it is sufficient to consider the Lorentz-violating linearized Ginzburg-Landau equation at the threshold of superconductivity, Eq. \eqref{eqdifescalarcomplexequatio} takes the form of
\begin{eqnarray}\label{tipo2superconductglmodifiedwconst}
 \Bigl(\mathbf{\nabla}^2 -\mu^2+2iexB\partial_y-2^2x^2B^2+2igWB\partial_z\Bigr)\, \phi =0,  
\end{eqnarray} where $\partial_a = \partial/\partial a $. Due to the boundary conditions on the surface and interior of the superconductor, which were mentioned earlier, we are led to consider solutions of the form
\begin{eqnarray}\label{tipo2solucmodifiedgl}
    \phi(\mathbf{x}) = \varphi(x) \, e^{i(k_yy+k_zz)}.
\end{eqnarray} Substituting Eq. \eqref{tipo2solucmodifiedgl} into Eq. \eqref{tipo2superconductglmodifiedwconst}, we obtain
\begin{eqnarray}\label{eqschrodingertipo2superconductm12}
   \Biggl[-\frac{d^2}{dX^2} + (eBX)^2 - (|\mu^2|-k_z^2-2gWBk_z) \Biggr]\, \varphi = 0,
\end{eqnarray} where we have made the displacement $x=X-(k_y/eB)$. Eq. \eqref{eqschrodingertipo2superconductm12} can then be identified as the Schrödinger equation for a harmonic oscillator of mass $m=1/2$, frequency $\omega = 2eB$, and energy eigenvalue $E = |\mu^2|-k_z^2-2gWBk_z$. As we know, the solution to this equation is given in terms of Hermite polynomials, which are stationary and normalized solutions. However, there exists a limiting value of $B$ for which this type of solution can exist. This value, which is related to the maximum value that the potential $V(X) = (1/4)\omega^2X^2$ can have for the Schrödinger equation to have a solution, occurs when the energy eigenvalue is at its minimum, i.e., $ E_0 = \omega/2$. Thus, we have
\begin{eqnarray}
    B_{\text{max}} = \mu_0H_{\text{max}} = \frac{|\mu^2|-k_z^2}{e+2gWk_z}.
\end{eqnarray} The critical external magnetic field $H_{c2}$ is defined for the value of $k_z$ that makes the above expression a maximum, that is:
\begin{eqnarray}\label{hc2maximotipo2supercond}
    H_{c2}= \frac{2 e-2 \sqrt{e^2-4 |\mu^2| g^2  W^2}}{4 \mu_0 g^2 W^2}.
\end{eqnarray} Unlike what we find in equation \eqref{campmagcriticosupconduct1}, our result for the critical magnetic field $H_{c2}$, capable of destroying the superconductivity state of a type II superconductor, explicitly feels the effects of Lorentz violation characterized by the constant parameter $W$. Since the Lorentz-violating effects are expected to be very small, we can still approximate Eq. \eqref{hc2maximotipo2supercond} as \begin{eqnarray}
    H_{c2} \approx \frac{|\mu^2|}{\mu_0 e} + \frac{g^2 W^2 |\mu^2|^2}{\mu_0 e^3}.
\end{eqnarray} We can then identify the correction to the usual critical magnetic field as $\delta H_{c2} = g^2 W^2 |\mu^2|^2/\mu_0 e^3$. Note that this correction is intended to increase the value of $H_{c2}$, as we can see from Fig. \eqref{hc2xtgraph}. Also note that the linear behavior found in the usual case (black line) is in accordance with what is expected from the Ginzburg-Landau theory for superconductors in the vicinity of $T=T_c$.

\subsection{London's penetration depth}

In order to investigate the influence of Lorentz violation on the London's penetration depth, we must turn our attention to the current density found in Eq. \eqref{jvectorxgamma2violation}. Taking the curl of both sides of it and making use of Ampère-Maxwell’s law, we can write
\begin{eqnarray}\label{modifiedsecondlondonequation}
    \mathbf{\nabla}^2\mathbf{B} &=& \Gamma^2 \mathbf{B}+\frac{g}{e}\Gamma^2 \Biggl\{ 2W(\mathbf{x}) \bigl[ \mathbf{\nabla}\times\mathbf{B} \bigr] - 2\mathbf{B}\times \mathbf{\nabla} W(\mathbf{x}) \nonumber\\
    &+&  \bigl[\mathbf{A}\cdot\mathbf{\nabla}\bigr]\mathbf{\nabla}W(\mathbf{x}) - \bigl[\mathbf{\nabla}W(\mathbf{x})\cdot\mathbf{\nabla}\bigr]\mathbf{A} \nonumber\\
    &+& \mathbf{\nabla}W(\mathbf{x})\bigl[\mathbf{\nabla}\cdot\mathbf{A}\bigr] - \mathbf{A}\mathbf{\nabla}^2W(\mathbf{x})\Biggr\}.
\end{eqnarray} From the above equation, in the absence of any violation, we recover the result that defines the usual London penetration depth $\lambda_L$, namely
\begin{eqnarray}
    \mathbf{\nabla}^2\mathbf{B} = \frac{1}{\lambda_L^2}\, \mathbf{B}
\end{eqnarray} where $\lambda_L= \Gamma^{-1}= \sqrt{\lambda/e |\mu^2|}$. A much simpler form of Eq. \eqref{modifiedsecondlondonequation} is obtained when the violation parameter is a constant $W$:
\begin{eqnarray}
 \mathbf{\nabla}^2\mathbf{B} &=& \frac{1}{\lambda_L^2}\, \left[ \mathbf{B}+\left(\frac{2 g W}{e}\right) \, \mathbf{\nabla}\times\mathbf{B} \right]   
\end{eqnarray}

\section{Final Remarks}

In this paper we study the effects of non-minimal Lorentz-violation operators in superconductivity. By constructing a Lorentz-Violating Ginzburg-Landau theory of superconductivity with five-dimensional operators, we discuss the influence of a higher dimensional Lorentz-Violating operator in the London's depth penetration, in the coherence length and critical magnetic field that breaks the superconducting state. 

The higher-dimensional Lorentz-violating operator was included by considering a non-minimal coupling between the complex scalar field, which stands as the field for the Cooper pairs, and the gauge field. We shown that the coherence length is not affected by the Lorentz-violating contributions.

Also, for type I superconductors, we have demonstrated that the critical magnetic field is insensitive to the Lorentz-violating contributions. However, the critical magnetic field $H_{c2}$ for type II superconductors is affected by the Lorentz-violating contributions, so that the increasing in the LV term $gW$ leads to an increase in the value of $H_{c2}$.

The London's depth penetration is drastically modified by the presence of the Lorentz-violating contributions. Here we have considered two cases, namely, the constant $W$ parameter case and a position-dependent $W(\bf{x})$ case. For both cases we have depicted the behaviour of the magnetic field inside the superconductor in order to investigate the influence of the Lorentz-violating term in the Meissner effect.

Finally, it was evident that under certain conditions controlled by the Lorentz-violating contribution, there exists a first-order phase transition between the superconducting state and an insulating state. It is important to highlight here that the Lorentz-violating parameter $W$ plays the role of local impurities that could break the homogeneity and isotropy of the system as a whole. Also, it is worth to mention that the addition of new terms violating Lorentz symmetry, whether in the pure scalar sector \cite{Furtado:2020iom}, or through different couplings between the scalar sector and the gauge sector or even direct modifications in the pure gauge sector, as proposed by other types of electrodynamics, should lead to results completely different from those presented here. Such additional terms directly impact the equations for the scalar and gauge fields, from which the quantities calculated here are derived. Specifically, we expect modifications both in the critical magnetic field $H_{c2}$ for a type II superconductor and in the behavior of the magnetic field within the superconductor, i.e., related to the London penetration depth. In fact, this will be a subject of investigation in future works.

\acknowledgments

\hspace{0.5cm} JF would like to thank the Fundação Cearense de Apoio ao Desenvolvimento Cient\'{i}fico e Tecnol\'{o}gico (FUNCAP) under the grant PRONEM PNE0112-00085.01.00/16 for financial support and Gazi University for the kind hospitality.


\begin{thebibliography}{99}

\bibitem{Kos1}
V. A. Kostelecky and S. Samuel,
Phys.\ Rev.\ D {\bf 39}, 683 (1989)

\bibitem{Kos2}
V. A. Kostelecky and R. Potting,
Nucl.\ Phys.\ B {\bf 359}, 545 (1991)

\bibitem{Kos3}
V. A. Kostelecky and S. Samuel,
Phys.\ Rev.\ D {\bf 40}, 1886 (1989)

\bibitem{Kos4}
V. A. Kostelecky and S. Samuel,
Phys.\ Rev.\ Lett. {\bf 63}, 224 (1989)

\bibitem{Colladay:1996iz} 
  D.~Colladay and V.~A.~Kostelecky,
  %``CPT violation and the standard model,''
  Phys.\ Rev.\ D {\bf 55}, 6760 (1997)
  [hep-ph/9703464].
  
\bibitem{Colladay:1998fq} 
  D.~Colladay and V.~A.~Kostelecky,
  %``Lorentz violating extension of the standard model,''
  Phys.\ Rev.\ D {\bf 58}, 116002 (1998)
  [hep-ph/9809521].  
  
  \bibitem{Edwards:2018lsn}
  B.~R.~Edwards and V.~A.~Kostelecky,
  %``Riemann–Finsler geometry and Lorentz-violating scalar fields,''
  Phys.\ Lett.\ B {\bf 786} (2018) 319
  doi:10.1016/j.physletb.2018.10.011
 
  
\bibitem{Edwards:2019lfb}
  B.~R.~Edwards and V.~A.~Kostelecký,
  %``Searching for CPT Violation with Neutral-Meson Oscillations,''
  Phys.\ Lett.\ B {\bf 795} (2019) 620
  %doi:10.1016/j.physletb.2019.07.012
  %[arXiv:1907.05206 [hep-ph]].
  %%CITATION = doi:10.1016/j.physletb.2019.07.012;%%
  %4 citations counted in INSPIRE as of 19 Nov 2019

  %\cite{Jackiw:1999yp}
\bibitem{Jackiw:1999yp}
R.~Jackiw and V.~A.~Kostelecky,
%``Radiatively induced Lorentz and CPT violation in electrodynamics,''
Phys. Rev. Lett. \textbf{82}, 3572-3575 (1999)
%doi:10.1103/PhysRevLett.82.3572
%[arXiv:hep-ph/9901358 [hep-ph]].
%473 citations counted in INSPIRE as of 27 Nov 2023

%\cite{Kostelecky:2001jc}
\bibitem{Kostelecky:2001jc}
V.~A.~Kostelecky, C.~D.~Lane and A.~G.~M.~Pickering,
%``One loop renormalization of Lorentz violating electrodynamics,''
Phys. Rev. D \textbf{65}, 056006 (2002)
%doi:10.1103/PhysRevD.65.056006
%[arXiv:hep-th/0111123 [hep-th]].
%258 citations counted in INSPIRE as of 27 Nov 2023

%\cite{Ferrari:2021eam}
\bibitem{Ferrari:2021eam}
A.~F.~Ferrari, J.~Furtado, J.~F.~Assun\c{c}\~ao, T.~Mariz and A.~Y.~Petrov,
%``One-loop calculations in Lorentz-breaking theories and proper-time method,''
EPL \textbf{136}, no.2, 21002 (2021)
%doi:10.1209/0295-5075/ac419a
%[arXiv:2109.11901 [hep-th]].
%7 citations counted in INSPIRE as of 27 Nov 2023

%\cite{MINOS:2008fnv}
\bibitem{MINOS:2008fnv}
P.~Adamson \textit{et al.} [MINOS],
%``Testing Lorentz Invariance and CPT Conservation with NuMI Neutrinos in the MINOS Near Detector,''
Phys. Rev. Lett. \textbf{101}, 151601 (2008)
%doi:10.1103/PhysRevLett.101.151601
%[arXiv:0806.4945 [hep-ex]].
%130 citations counted in INSPIRE as of 27 Nov 2023

%\cite{MINOS:2010kat}
\bibitem{MINOS:2010kat}
P.~Adamson \textit{et al.} [MINOS],
%``A Search for Lorentz Invariance and CPT Violation with the MINOS Far Detector,''
Phys. Rev. Lett. \textbf{105}, 151601 (2010)
%doi:10.1103/PhysRevLett.105.151601
%[arXiv:1007.2791 [hep-ex]].
%118 citations counted in INSPIRE as of 27 Nov 2023

%\cite{MiniBooNE:2011pix}
\bibitem{MiniBooNE:2011pix}
A.~A.~Aguilar-Arevalo \textit{et al.} [MiniBooNE],
%``Test of Lorentz and CPT violation with Short Baseline Neutrino Oscillation Excesses,''
Phys. Lett. B \textbf{718}, 1303-1308 (2013)
%doi:10.1016/j.physletb.2012.12.020
%[arXiv:1109.3480 [hep-ex]].
%89 citations counted in INSPIRE as of 27 Nov 2023

%\cite{Furtado:2014cja}
\bibitem{Furtado:2014cja}
J.~Furtado and T.~Mariz,
%``Lorentz-violating Euler-Heisenberg effective action,''
Phys. Rev. D \textbf{89}, no.2, 025021 (2014)
%doi:10.1103/PhysRevD.89.025021
%[arXiv:1401.0492 [hep-ph]].
%10 citations counted in INSPIRE as of 27 Nov 2023

%\cite{Kostelecky:2016kkn}
\bibitem{Kostelecky:2016kkn}
V.~A.~Kosteleck\'y, A.~C.~Melissinos and M.~Mewes,
%``Searching for photon-sector Lorentz violation using gravitational-wave detectors,''
Phys. Lett. B \textbf{761}, 1-7 (2016)
doi:10.1016/j.physletb.2016.08.001
[arXiv:1608.02592 [gr-qc]].
%68 citations counted in INSPIRE as of 27 Nov 2023

%\cite{Assuncao:2018jkq}
\bibitem{Assuncao:2018jkq}
J.~F.~Assun\c{c}\~ao, T.~Mariz, J.~R.~Nascimento and A.~Y.~Petrov,
%``Induced Chern-Simons modified gravity at finite temperature,''
JHEP \textbf{08}, 072 (2018)
%doi:10.1007/JHEP08(2018)072
%[arXiv:1805.11049 [hep-th]].
%8 citations counted in INSPIRE as of 27 Nov 2023

%\cite{Leite:2013pca}
\bibitem{Leite:2013pca}
J.~Leite, T.~Mariz and W.~Serafim,
%``The induced higher derivative Lorentz-violating Chern\textendash{}Simons term at finite temperature,''
J. Phys. G \textbf{40}, 075003 (2013)
%doi:10.1088/0954-3899/40/7/075003
%[arXiv:1712.09675 [hep-th]].
%22 citations counted in INSPIRE as of 27 Nov 2023

%\cite{Leite:2011jg}
\bibitem{Leite:2011jg}
J.~Leite and T.~Mariz,
%``Induced Lorentz-violating terms at finite temperature,''
EPL \textbf{99}, no.2, 21003 (2012)
doi:10.1209/0295-5075/99/21003
[arXiv:1110.2127 [hep-th]].
%18 citations counted in INSPIRE as of 27 Nov 2023

%\cite{Assuncao:2016fko}
\bibitem{Assuncao:2016fko}
J.~F.~Assuncao, T.~Mariz and A.~Y.~Petrov,
%``Nonanalyticity of the induced Carroll-Field-Jackiw term at finite temperature,''
EPL \textbf{116}, no.3, 31003 (2016)
doi:10.1209/0295-5075/116/31003
[arXiv:1609.09120 [hep-th]].
%5 citations counted in INSPIRE as of 27 Nov 2023

%\cite{Araujo:2023izx}
\bibitem{Araujo:2023izx}
M.~C.~Ara\'ujo, J.~Furtado and R.~V.~Maluf,
%``Lorentz-violating extension of scalar QED at finite temperature,''
Phys. Lett. B \textbf{844}, 138064 (2023)
doi:10.1016/j.physletb.2023.138064
[arXiv:2306.06959 [hep-th]].
%1 citations counted in INSPIRE as of 27 Nov 2023

%\cite{Mariz:2022oib}
\bibitem{Mariz:2022oib}
T.~Mariz, J.~R.~Nascimento and A.~Petrov,
%``Lorentz Symmetry Breaking\textemdash{}Classical and Quantum Aspects,''
Springer, 2023,
ISBN 978-3-031-20119-6, 978-3-031-20120-2
%doi:10.1007/978-3-031-20120-2
%[arXiv:2205.02594 [hep-th]].
%14 citations counted in INSPIRE as of 27 Nov 2023

%\cite{Myers:2003fd}
\bibitem{Myers:2003fd}
R.~C.~Myers and M.~Pospelov,
%``Ultraviolet modifications of dispersion relations in effective field theory,''
Phys. Rev. Lett. \textbf{90}, 211601 (2003)
%doi:10.1103/PhysRevLett.90.211601
%[arXiv:hep-ph/0301124 [hep-ph]].
%502 citations counted in INSPIRE as of 27 Nov 2023


%\cite{Gomes:2009ch}
\bibitem{Gomes:2009ch}
M.~Gomes, J.~R.~Nascimento, A.~Y.~Petrov and A.~J.~da Silva,
%``On the aether-like Lorentz-breaking actions,''
Phys. Rev. D \textbf{81}, 045018 (2010)
%doi:10.1103/PhysRevD.81.045018
%[arXiv:0911.3548 [hep-th]].
%150 citations counted in INSPIRE as of 02 Jan 2024


%\cite{Gazzola:2010vr}
\bibitem{Gazzola:2010vr}
G.~Gazzola, H.~G.~Fargnoli, A.~P.~Baeta Scarpelli, M.~Sampaio and M.~C.~Nemes,
%``QED with minimal and nonminimal couplings: on the quantum generation of Lorentz violating terms in the pure photon sector,''
J. Phys. G \textbf{39}, 035002 (2012)
%doi:10.1088/0954-3899/39/3/035002
%[arXiv:1012.3291 [hep-th]].
%89 citations counted in INSPIRE as of 02 Jan 2024

%\cite{Mariz:2011ed}
\bibitem{Mariz:2011ed}
T.~Mariz, J.~R.~Nascimento and A.~Y.~Petrov,
%``On the perturbative generation of the higher-derivative Lorentz-breaking terms,''
Phys. Rev. D \textbf{85}, 125003 (2012)
%doi:10.1103/PhysRevD.85.125003
%[arXiv:1111.0198 [hep-th]].
%67 citations counted in INSPIRE as of 02 Jan 2024

%\cite{BaetaScarpelli:2013rmt}
\bibitem{BaetaScarpelli:2013rmt}
A.~P.~Baeta Scarpelli, T.~Mariz, J.~R.~Nascimento and A.~Y.~Petrov,
%``Four-dimensional aether-like Lorentz-breaking QED revisited and problem of ambiguities,''
Eur. Phys. J. C \textbf{73}, 2526 (2013)
%doi:10.1140/epjc/s10052-013-2526-3
%[arXiv:1304.2256 [hep-th]].
%56 citations counted in INSPIRE as of 02 Jan 2024

%\cite{Kostelecky:2018yfa}
\bibitem{Kostelecky:2018yfa}
V.~A.~Kosteleck\'y and Z.~Li,
%``Gauge field theories with Lorentz-violating operators of arbitrary dimension,''
Phys. Rev. D \textbf{99}, no.5, 056016 (2019)
%doi:10.1103/PhysRevD.99.056016
%[arXiv:1812.11672 [hep-ph]].
%86 citations counted in INSPIRE as of 02 Jan 2024

%\cite{Kostelecky:2011gq}
\bibitem{Kostelecky:2011gq}
A.~Kostelecky and M.~Mewes,
%``Neutrinos with Lorentz-violating operators of arbitrary dimension,''
Phys. Rev. D \textbf{85}, 096005 (2012)
%doi:10.1103/PhysRevD.85.096005
%[arXiv:1112.6395 [hep-ph]].
%312 citations counted in INSPIRE as of 27 Nov 2023

%\cite{Mariz:2010fm}
\bibitem{Mariz:2010fm}
T.~Mariz,
%``Radiatively induced Lorentz-violating operator of mass dimension five in QED,''
Phys. Rev. D \textbf{83}, 045018 (2011)
%doi:10.1103/PhysRevD.83.045018
%[arXiv:1010.5013 [hep-th]].
%55 citations counted in INSPIRE as of 27 Nov 2023

%\cite{Rubtsov:2012kb}
\bibitem{Rubtsov:2012kb}
G.~Rubtsov, P.~Satunin and S.~Sibiryakov,
%``On calculation of cross sections in Lorentz violating theories,''
Phys. Rev. D \textbf{86}, 085012 (2012)
%doi:10.1103/PhysRevD.86.085012
%[arXiv:1204.5782 [hep-ph]].
%39 citations counted in INSPIRE as of 27 Nov 2023

%\cite{Kostelecky:2009zp}
\bibitem{Kostelecky:2009zp}
V.~A.~Kostelecky and M.~Mewes,
%``Electrodynamics with Lorentz-violating operators of arbitrary dimension,''
Phys. Rev. D \textbf{80}, 015020 (2009)
%doi:10.1103/PhysRevD.80.015020
%[arXiv:0905.0031 [hep-ph]].
%485 citations counted in INSPIRE as of 02 Jan 2024
  
\bibitem{Assuncao:2015lfa}
J.~Assunção and T.~Mariz,
%``Radiatively induced CPT-odd Chern-Simons term in massless QED,''
EPL \textbf{110}, no.4, 41002 (2015)
%[arXiv:1505.08156 [hep-th]].

\bibitem{Gos1}
P.~Goswami and S.~Tewari,
%``Axionic field theory of (3+1)-dimensional Weyl semimetals,''
Phys. Rev. B \textbf{88} (2013) no.24, 245107
%[arXiv:1210.6352 [cond-mat.mes-hall]].

\bibitem{Gos2}
P.~Goswami, G.~Sharma and S.~Tewari,
%``Optical activity as a test for dynamic chiral magnetic effect of Weyl semimetals,''
Phys. Rev. B \textbf{92} (2015) no.16, 161110
%[arXiv:1404.2927 [cond-mat.str-el]].
%28 citations counted in INSPIRE as of 20 May 2020

\bibitem{Gos3}
G.~Sharma, P.~Goswami and S.~Tewari,
%``Chiral anomaly and longitudinal magnetotransport in type-II Weyl semimetals,''
Phys. Rev. B \textbf{96} (2017) no.4, 045112
%[arXiv:1608.06625 [cond-mat.mes-hall]].
%19 citations counted in INSPIRE as of 20 May 2020

\bibitem{Bazeia:2016pra}
D.~Bazeia, F.~Brito and J.~Mota-Silva,
%``Kondo effect from a Lorentz-violating domain wall description of superconductivity,''
Phys. Lett. B \textbf{762} (2016), 327-333
%[arXiv:1606.07051 [hep-th]].

%\cite{Furtado:2020iom}
\bibitem{Furtado:2020iom}
J.~Furtado, R.~M.~M.~Costa Filho, A.~F.~Morais and I.~C.~Jardim,
%``Effects of Lorentz violation in superconductivity,''
EPL \textbf{136}, no.5, 51001 (2021)
%doi:10.1209/0295-5075/ac36f0
%[arXiv:2009.02301 [cond-mat.supr-con]].
%2 citations counted in INSPIRE as of 25 Dec 2023

%\cite{Granado:2023czl}
\bibitem{Granado:2023czl}
D.~R.~Granado,
%``Lorentz violation emergence in a superconductivity phase transition scenario,''
EPL \textbf{142}, no.5, 54001 (2023)
%doi:10.1209/0295-5075/acd8ea
%0 citations counted in INSPIRE as of 25 Dec 2023

\bibitem{Katsnelson:2012cz}
M.~Katsnelson and G.~Volovik,
%``Quantum electrodynamics with anisotropic scaling: Heisenberg-Euler action and Schwinger pair production in the bilayer graphene,''
JETP Lett. \textbf{95} (2012), 411-415
%[arXiv:1203.1578 [cond-mat.str-el]].

\bibitem{Baym:2020uos}
G.~Baym, D.~Beck, J.~P.~Filippini, C.~Pethick and J.~Shelton,
%``Searching for low mass dark matter via phonon creation in superfluid 4He,''
%[arXiv:2005.08824 [hep-ph]].


\bibitem{Pereira:2009vb}
E.~R.~Pereira and F.~Moraes,
%``Flowing Liquid Crystal Simulating the Schwarzschild Metric,''
Central Eur. J. Phys. \textbf{9} (2011), 1100-1105
%[arXiv:0910.1314 [gr-qc]].

\bibitem{Ginzburg:1950sr}
V.~Ginzburg and L.~Landau,
%``On the Theory of superconductivity,''
Zh. Eksp. Teor. Fiz. \textbf{20} (1950), 1064-1082

\bibitem{Bardeen:1957mv}
J.~Bardeen, L.~Cooper and J.~Schrieffer,
%``Theory of superconductivity,''
Phys. Rev. \textbf{108} (1957), 1175-1204

\bibitem{Zhou:2015qka}
T.~Zhou, Y.~Gao and Z.~Wang,
%``Superconductivity in doped inversion-symmetric Weyl semimetals,''
Phys. Rev. B \textbf{93} (2016) no.9, 094517
%[arXiv:1510.01051 [cond-mat.supr-con]].

\bibitem{Bednik:2015tha}
G.~Bednik, A.~Zyuzin and A.~Burkov,
%``Superconductivity in Weyl metals,''
Phys. Rev. B \textbf{92} (2015) no.3, 035153
%[arXiv:1506.05109 [cond-mat.str-el]].

\bibitem{Wei:2014vsa}
H.~Wei, S.~P.~Chao and V.~Aji,
%``Odd-parity superconductivity in Weyl semimetals,''
Phys. Rev. B \textbf{89} (2014) no.1, 014506
%[arXiv:1305.7233 [cond-mat.supr-con]].

\bibitem{Roy:2013aya}
B.~Roy and V.~Juricic,
%``Strain-induced time-reversal odd superconductivity in graphene,''
Phys. Rev. B \textbf{90} (2014) no.4, 041413
%[arXiv:1309.0507 [cond-mat.mes-hall]].

\bibitem{Cohnitz:2017vsr}
L.~Cohnitz, A.~De Martino, W.~Häusler and R.~Egger,
%``Proximity-induced superconductivity in Landau-quantized graphene monolayers,''
Phys. Rev. B \textbf{96} (2017) no.14, 140506
%[arXiv:1708.02892 [cond-mat.mes-hall]].
%0 citations counted in INSPIRE as of 21 May 2020

\bibitem{Madsen:1999ci}
J.~Madsen,
%``Probing strange stars and color superconductivity by r-mode instabilities in millisecond pulsars,''
Phys. Rev. Lett. \textbf{85} (2000) no.1, 10-13
%[arXiv:astro-ph/9912418 [astro-ph]].

\bibitem{Bonanno:2011ch}
L.~Bonanno and A.~Sedrakian,
%``Composition and stability of hybrid stars with hyperons and quark color-superconductivity,''
Astron. Astrophys. \textbf{539} (2012), A16
%[arXiv:1108.0559 [astro-ph.SR]].

\bibitem{Ebert:2007ey}
D.~Ebert, A.~Tyukov and V.~Zhukovsky,
%``Color superconductivity in the static Einstein Universe,''
Phys. Rev. D \textbf{76} (2007), 064029
%[arXiv:hep-th/0703213 [hep-th]].

\bibitem{Gao:2012aw}
X.~Gao, A.~M.~Garcia-Garcia, H.~B.~Zeng and H.~Q.~Zhang,
%``Normal modes and time evolution of a holographic superconductor after a quantum quench,''
JHEP \textbf{06} (2014), 019
%[arXiv:1212.1049 [hep-th]].

\bibitem{Herzog:2009xv}
C.~P.~Herzog,
%``Lectures on Holographic Superfluidity and Superconductivity,''
J. Phys. A \textbf{42} (2009), 343001
%[arXiv:0904.1975 [hep-th]].

\bibitem{Ghoroku:2019trx}
K.~Ghoroku, K.~Kashiwa, Y.~Nakano, M.~Tachibana and F.~Toyoda,
%``Color superconductivity in a holographic model,''
Phys. Rev. D \textbf{99} (2019) no.10, 106011
%[arXiv:1902.01093 [hep-th]].

%\cite{Nobre:1999mj}
\bibitem{Nobre:1999mj}
F.~A.~S.~Nobre and C.~A.~S.~Almeida,
%``Pauli's term and fractional spin,''
Phys. Lett. B \textbf{455} (1999), 213-216
doi:10.1016/S0370-2693(99)00475-X
[arXiv:hep-th/9904159 [hep-th]].
%25 citations counted in INSPIRE as of 26 Jun 2024

%\cite{Furtado:2011zz}
\bibitem{Furtado:2011zz}
J.~S.~N.~Furtado and F.~A.~S.~Nobre,
%``The Pauli term as a generator of fractional spin,''
Mod. Phys. Lett. A \textbf{26} (2011), 1427-1432
doi:10.1142/S0217732311035912
%0 citations counted in INSPIRE as of 27 Jun 2024

%\cite{Semenoff:1988jr}
\bibitem{Semenoff:1988jr}
G.~W.~Semenoff,
%``Canonical Quantum Field Theory with Exotic Statistics,''
Phys. Rev. Lett. \textbf{61} (1988), 517
doi:10.1103/PhysRevLett.61.517
%286 citations counted in INSPIRE as of 26 Jun 2024

%\cite{Dunne:1994uy}
\bibitem{Dunne:1994uy}
G.~V.~Dunne,
%``Selfdual Chern-Simons theories,''
[arXiv:hep-th/9410065 [hep-th]].
%23 citations counted in INSPIRE as of 26 Jun 2024

%\cite{Wilczek:1982wy}
\bibitem{Wilczek:1982wy}
F.~Wilczek,
%``Quantum Mechanics of Fractional Spin Particles,''
Phys. Rev. Lett. \textbf{49} (1982), 957-959
doi:10.1103/PhysRevLett.49.957
%1253 citations counted in INSPIRE as of 26 Jun 2024

%\cite{Furtado:2020olp}
\bibitem{Furtado:2020olp}
J.~Furtado, A.~C.~A.~Ramos and J.~F.~Assun\c{c}\~ao,
%``Effects of Lorentz violation in the Bose-Einstein condensation,''
EPL \textbf{132} (2020) no.3, 31001
%doi:10.1209/0295-5075/132/31001
%[arXiv:2009.07034 [hep-th]].
%1 citations counted in INSPIRE as of 26 Apr 2021

%\cite{Casana:2011bv}
\bibitem{Casana:2011bv}
R.~Casana and K.~A.~T.~da Silva,
%``Lorentz-violating effects in the Bose\textendash{}Einstein condensation of an ideal bosonic gas,''
Mod. Phys. Lett. A \textbf{30} (2015) no.07, 1550037
%doi:10.1142/S0217732315500376
%[arXiv:1106.5534 [hep-th]].
%16 citations counted in INSPIRE as of 26 Apr 2021
    

     

\end{thebibliography}
\end{document}